\documentclass[11pt]{article}
\usepackage{jheppub}
\usepackage{amsmath,amssymb,amsfonts,graphicx,slashed,color, mathtools, amsthm}
\usepackage{comment}
\usepackage{enumerate}
\usepackage{float}
\usepackage{caption}
\usepackage{subcaption}
\usepackage{pdflscape}

\newcommand{\be}{\begin{equation}}
\newcommand{\ee}{\end{equation}}
\newcommand{\bea}{\begin{eqnarray}}
\newcommand{\eea}{\end{eqnarray}}
\newcommand{\beas}{\begin{eqnarray*}}
\newcommand{\eeas}{\end{eqnarray*}}
\newcommand{\ba}{\begin{array}}
\newcommand{\ea}{\end{array}}

\title{Suggestions of decreasing dark energy from supernova and BAO data }

\author[1]{Mark Van Raamsdonk,}
\author[1, 2, 3]{Chris Waddell}

\affiliation[1]{Department of Physics and Astronomy, University of British Columbia,\\
6224 Agricultural Road, Vancouver, B.C.\ V6T 1Z1, Canada}
\affiliation[2]{Kavli Institute for Theoretical Physics, University of California, Santa Barbara, CA 93106, USA}
\affiliation[3]{Perimeter Institute for Theoretical Physics, 31 Caroline St N, Waterloo, ON\ N2L 2Y5, Canada.}

\emailAdd{mav@phas.ubc.ca}
\emailAdd{cwaddell@perimeterinstitute.ca}

\abstract{The potential energy from a time-dependent scalar field provides a possible explanation for the observed cosmic acceleration. In this paper, we investigate how data from supernova and bary acoustic oscillation surveys constrain the possible evolution of a single scalar field over the period of time (roughly half the age of the universe) for which these data are available. Taking a linear approximation to the scalar potential $V(\phi) = V_0 + V_1 \phi$ around the present value, a likelihood analysis appears to significantly prefer models with a decreasing potential energy at present, with approximately $99.99 \%$ of the $\exp(-\chi^2/2)$ distribution having $V_1 > 0$ in a convention where $\dot{\phi} \le 0$ at present. The models favoured by the distribution typically have an order one decrease $\langle |{\rm Range}[V(\phi(t))] / V(t_0)| \rangle \approx 0.36$ in the scalar potential energy over the time frame corresponding to $z < 2$. According to the likelihood analysis, the $\Lambda$CDM model with no variation in dark energy appears to be significantly disfavoured in the context of the linear potential model, but this should be interpreted cautiously since model selection criteria that make use of $\Delta \chi^2$ while ignoring parameter space volumes still favour $\Lambda$CDM. Working with a second order approximation to the potential, the supernova data can be fit well for a wide range of possible potentials, including models where the universe has already stopped accelerating. }

\begin{document}

\maketitle

\section{Introduction}

Understanding the origin of the observed accelerated expansion of the universe \cite{perlmutter1999measurements, riess1998observational} is a key question in theoretical cosmology. From the effective field theory point of view, the simplest explanation is probably a positive cosmological constant. However, effective theories with a positive cosmological constant have been difficult to realize microscopically, at least in the context of string theory and/or holography \cite{Obied:2018sgi,Danielsson:2018ztv,Bena:2023sks}, though there are a number of interesting proposals \cite{Banks:2001px,Strominger:2001pn,Alishahiha:2004md,Gorbenko:2018oov,Coleman:2021nor,Freivogel2005,McFadden:2009fg,Banerjee:2018qey,Susskind:2021dfc}. There also appears to be a tension presently with fitting both supernova redshift vs brightness data and CMB observations with a single set of model parameters in the simple $\Lambda$CDM model with a positive cosmological constant \cite{DiValentino:2021izs}. The accelerated expansion may be explained more generally by the time-dependent potential energy of a rolling scalar field that is presently at a positive value of its potential  \cite{Peebles:1987ek,Ratra:1987rm,Caldwell:1997ii,Dutta:2018vmq,Visinelli:2019qqu,sen2021cosmological}.\footnote{Strictly speaking, it is only required to have been at positive values in the relatively recent past; we will find examples where it has already descended to a negative value.} In \cite{VanRaamsdonk:2022rts} (see also \cite{Antonini:2022blk}), we have argued that such models with scalars that vary on cosmological time scales are a generic expectation from the point of view of the holographic approach to quantum gravity. Further, in the scenario with time-dependent scalars rolling towards a negative extremum of the potential, holography might be used to give a fully microscopic description of the cosmological physics \cite{Maldacena:2004rf, McInnes:2004nx,Cooper:2018cmb, Antonini2019,VanRaamsdonk:2021qgv, Antonini:2022blk, Antonini:2022xzo}.

Motivated by these theoretical considerations, we would like to understand how strongly scalar field evolution is constrained by observation. Is there room for models with significant variation in the scalar potential energy in our recent cosmological history (e.g. for $z \lesssim 2$ where supernova data are available), or do observations already imply that we are very close to a $\Lambda$CDM model with constant dark energy density? More generally, assuming a single field scalar field model, how much can we learn about the potential through the most direct observations of the scale factor evolution? In this work, we access this scale factor evolution observationally by making use of the redshift vs brightness data in Pantheon+ \cite{Scolnic:2021amr}, the most up-to-date and comprehensive publicly available catalogue of type Ia supernovae, as well as  baryon acoustic oscillation (BAO) data from various surveys \cite{Ross:2014qpa, BOSS:2016wmc, eBOSS:2020lta, eBOSS:2020hur, eBOSS:2020fvk, eBOSS:2020uxp, eBOSS:2020gbb, eBOSS:2020tmo}.  Related analyses based on older data sets may be found in \cite{Kallosh:2003bq, wang2004current, Perivolaropoulos:2004yr, Avelino:2004vy, Sahlen:2006dn, Huterer:2006mv}.

\begin{figure}
    \centering
    \includegraphics[scale=0.5]{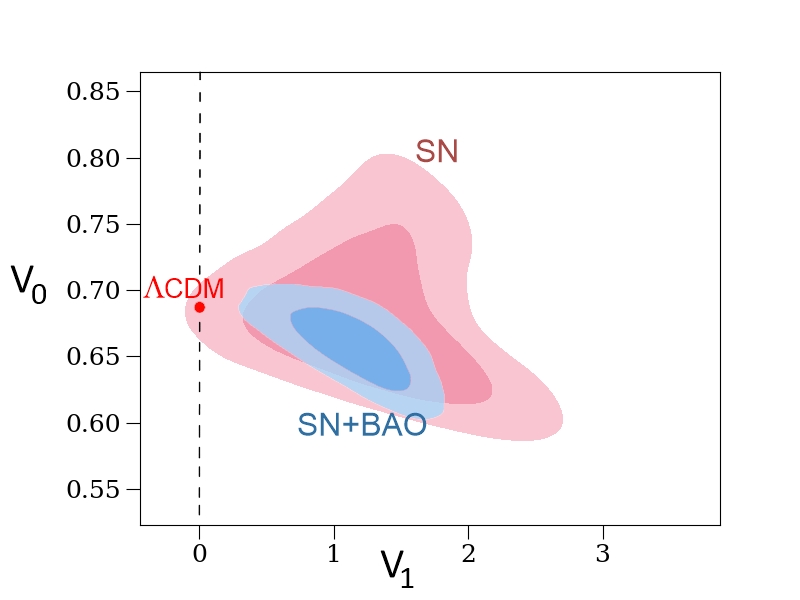}
    \caption{Distribution of parameter values $V_0$ and $V_1$ in a linear approximation $V(\phi) = V_0 + V_1 \phi$ to the scalar potential in the convention that $\phi=0, \dot{\phi} \le 0$ presently. The $\exp(-\chi^2/2)$ likelihood function is computed by comparing model predictions to redshift vs brightness data in the Pantheon+ supernova catalogue (68\% and 95\% contours shown in pink) and to combined supernova and BAO data (blue). Best fit $\Lambda$CDM model is indicated in red. Positive $V_1$ (corresponding to decreasing potential energy at present in our conventions) accounts for $\approx 99\%$ of the distribution with only supernova data and $\approx 99.99 \%$ with supernova and BAO data. Here, $V$ is normalized so that $V_0 = \Omega_\Lambda$ in a $\Lambda$CDM model with $V_1 = 0$, and our conventions for $V_1$ are such that an order one value typically corresponds to an order one decrease in $V$ over a Hubble time.}
    \label{fig:V0V1}
\end{figure}

We begin in Section \ref{sec:theory} with some theory. Starting from the Friedmann equation and scalar field evolution equation, we show how the time-dependence of scalar field potential and kinetic energies can be deduced from knowledge of the scale factor evolution together with the present densities of matter and radiation. For a single scalar field model in the approximation where radiation can be neglected, the evolution $a(t)$ together with $\Omega_M$ are enough to deduce both the scalar potential $V(\phi)$ and the time-dependence of the scalar field. Thus, assuming only a single scalar field, we in principle have direct access to its potential via observations of $a(t)$ and $\Omega_M$. 

In practice, $\Omega_M$ is only loosely constrained by direct observations so even with precise knowledge of the scale factor evolution, we could only pin down the potential to within a one-parameter family. Even fixing $\Omega_M$, we find that small changes in the scale factor can correspond to significant changes in the potential. As a result, we will find that the modest uncertainties present in the supernova/BAO data and the associated scale factor evolution lead to much larger uncertainties in the deduced scalar potential, and a significant range of potential parameters are possible. We will find that typical models that provide a good fit to the data have significant scalar field evolution and ${\cal O}(1)$ changes in the scalar potential $V(\phi)$ during the period $z \lesssim 2$ corresponding to the supernova data. 

In Section \ref{sec:observational}, we describe the procedure to constrain model parameters based on the supernova and BAO data. For a given scalar potential $V(\phi)$, the potential parameters together with $\Omega_M$ and $H_0$ determine a scale factor evolution $a(t)$ and thus a theoretical redshift vs brightness curve\footnote{Here, the mean absolute magnitude of the supernovae is an additional parameter that we allow to vary.} that can be compared with supernova data. For each set of model parameters, we can calculate a chi-squared value, following precisely the same methodology as the Pantheon+ cosmology analysis in \cite{brout2022pantheon+}. The chi-squared values are used to define a likelihood function (a relative probability for the parameter values) and we use a Markov Chain Monte Carlo (MCMC) algorithm to sample from the resulting probability distribution, yielding histograms and correlation plots for the various parameters. As a check, we have verified that our code accurately reproduces the distribution obtained in \cite{brout2022pantheon+} for the $w$CDM model. We also describe how to incorporate BAO data into the analysis, by adapting existing likelihoods for the relevant surveys to our scalar dark energy models. 

In Section \ref{sec:results}, we present our results for parameter constraints in time-dependent scalar field models. We begin in Section \ref{sec:results_linear} by considering supernova data only, and models with a linear potential $V(\phi) = V_0 + V_1 \phi$, which we can think of as the first-order Taylor series approximation to a more general potential expanded about the present value of the scalar field. Such an approximation will always be valid over sufficiently short time scales, though not necessarily over the range of times corresponding to the supernova data. Our results for the parameter constraints in this class of models are displayed in Figures \ref{fig:V01} and \ref{fig:OMplusV0vsV1}. We find that a large majority of the models providing a good fit to the data have a scalar potential that is presently decreasing with time. According to the $e^{- \chi^2/2}$ likelihood, $\approx 99.1\%$ of the distribution
has $V_1 > 0$ ($\phi$ is presently decreasing with time according to our conventions), with $V_1 \in [0.72, 1.77]$ accounting for 68\%.\footnote{Here $V$ is normalized so that $V_0 = \Omega_\Lambda$ in a $\Lambda$CDM model, and we are taking units where $8 \pi G / 3 = 1$. With our normalization of $V_1$, an order one value of $V_1$ results in a order one fractional change in the scalar potential over a Hubble time.} The scalar potential typically changes by a significant amount during the recent evolution, with $\langle |{\rm Range}[V(\phi(t))] / V(t_0)| \rangle \approx 0.97$ over the time scale from which data are available ($z \lesssim 2$). 
A final notable feature of the distribution is that it has significant support over all possible positive values of $\Omega_M$ up to about 0.35, though this range narrows considerably when taking into account BAO data.

In Section \ref{sec:results_linear_bao}, we constrain the parameters of the linear potential model further using both supernova and BAO data. As shown in Figure \ref{fig:V0V1}, this narrows the distribution to one that is almost entirely supported in the $V_1 > 0$ region, such that $\Lambda$CDM sits well outside the 95\% confidence region. In the marginalized distribution for the slope parameter $V_1$, we find $V_1 = 1.13 \pm 0.30$ and estimate that only  $\sim 0.01 \%$ of the distribution has $V_1 \le 0$. Thus, within the context of the linear potential models, our likelihood analysis disfavours the $\Lambda$CDM model at a level that would correspond to approximately $3.7\sigma$ for a Gaussian distribution. However, the distribution we find has significant non-Gaussianities, and we find that the $\Delta \chi^2$ between the best fit $\Lambda$CDM model and the best fit linear potential model is significantly smaller than would be expected at this significance level from a Gaussian distribution. As a result, model selection criteria that take into account only $\Delta \chi^2$ without considering parameter space volumes do not favour the linear potential models; for example, the two models are nearly degenerate according to the Akaike Information Criterion. The apparent discrepancy in significance between the likelihood analysis and the other model selection criteria arises from the fact that a relatively tiny volume of parameter space near the best fit $\Lambda$CDM model has a comparable $\chi^2$ to that model, while a much larger volume of parameter space near the global best fit has comparable $\chi^2$ to that model.

Results for the other parameters are shown in Figure \ref{fig:LinearPotential_Results_SNBAO} and Table \ref{tab:params_BAO}. Due to a reduction in the value and range of $V_{1}$ compared to the supernova-only analysis, the typical range of scalar field values during the recent evolution is reduced to $\langle |{\rm Range}[V(\phi(t))] / V(t_0)| \rangle \approx 0.36$. Compared to the supernova-only results there is a substantial narrowing of the distribution for $\Omega_{M}$, which is now constrained to the range $[0.25,0.33]$ with 95\% confidence. In these models, the scalar potential energy will become negative at some time in the future; we find that the median time for this is $\sim 2.1$ Hubble times or $\sim 29$ billion years.\footnote{See \cite{andrei2022rapidly} for another recent work suggesting that models descending to a negative scalar potential in the relatively near future can be consistent with data.}  

As a further exploration, in Section \ref{sec:results_quadratic}, we extend the analysis (of supernova data only) to include a quadratic term in the potential, which now takes the form $V(\phi) = V_{0} + V_{1} \phi + \frac{1}{2} V_{2} \phi^{2}$. Here, we find a very large range of possible potentials that can provide a good fit to the data. In particular, essentially any value of the quadratic coefficient $V_2$ is allowed. The wide range of possible potentials can be understood based on certain approximate degeneracies in the parameter space. For large positive values of $V_2$, the scalar field can oscillate about the minimum of the potential. For $\sqrt{V_2} \gg H_0$, the scalar field evolution is well-approximated by a WKB solution, and the evolution of the energy density of such a solution in the Friedmann equation is the same as for non-relativistic matter \cite{turner1983coherent}. Thus, we have an approximate degeneracy where a certain amount of matter is replaced by oscillating scalar field. For negative $V_2$, the scalar can spend a significant fraction of the evolution time close to the maximum of the parabolic potential, deviating from this value significantly only for early and/or late times. Notably, certain allowed models of this type have a significant late-time decrease in dark energy, such that the universe has already stopped accelerating. Examples of very different scalar potential evolution  that all provide a good fit to the data are shown in Figure \ref{fig:Vexamples}. 

In summary, taking into account only the most direct measurements of the scale factor evolution, a rather significant variation of the scalar potential in the recent history of the universe appears to be allowed and within the approximation of a linear potential, a scalar potential energy that is presently decreasing appears to be strongly preferred.  It will be very interesting to see how the parameters will be further constrained by newer data sets for supernovae and BAO as they become available. 

We note that the very recent DESI analysis \cite{DESI:2024mwx} that appeared after the original version of this paper also finds evidence for time-varying dark energy based on supernova and BAO data, though as we explain in the discussion, our analysis here explores a rather different (and perhaps more physical) slice through the space of cosmological models than than the $w_0 w_a$ models considered in \cite{DESI:2024mwx}.

\section{Theory} \label{sec:theory}

In this section, we review some background material on cosmology with rolling scalar fields. Generally, the scalar potential together with the current densities of matter and radiation and the current Hubble constant determine the scale factor evolution $a(t)$ and the scalar field evolution $\phi(t)$ through the coupled Friedmann equation and scalar evolution equation. For a single scalar field model, we can also go the other way and deduce the scalar potential from the observed scale factor evolution and present matter/radiation densities.

\subsection{Scalar potential from $a(t)$ and $\Omega_M$}
\label{sec:general}

Throughout our discussion we will assume a spatially flat FRW universe, with metric
\be
ds^2 = -dt^2 + a(t)^2 d \vec{x}^2 \: ,
\ee
and take the convention that the scale factor is $a=1$ presently. We focus on a model with a single scalar field $\phi$ with potential $V(\phi)$.\footnote{More generally, we could consider multiple scalar fields and a general metric on the scalar field target space.} Assuming that the scalar is minimally coupled to gravity (without curvature couplings), the evolution equation for the scalar field  (assumed to be spatially constant) is
\be
\label{scalarEOM}
\ddot{\phi} + 3 H \dot{\phi} +  V'(\phi) = 0 \: ,
\ee
where a dot denotes a $t$ derivative and $H = \dot a/a$ is the Hubble parameter. This is equivalent to the damped motion of a particle with position 
$\phi$ in a potential 
$V(\phi)$ with time-dependent damping parameter $3H(t)$. The gravitational equations yield the Friedmann equation
\be
\label{Friedmann}
H^2 = {8 \pi G \over 3}\left[\tilde{\rho} + {1 \over 2} \dot{\phi}^2 + V(\phi) \right]
\ee
where we define $\tilde{\rho}$ to be the energy density excluding that associated with the scalar field.

Taking the time derivative of the Friedmann equation and eliminating $\ddot{\phi}$ using the scalar equation gives an expression for the evolution of the scalar kinetic energy,
\be
\label{sc1}
{1 \over 2} \dot{\phi}^2 = -{1 \over 8 \pi G} \dot{H} + {1 \over 6} a {d \tilde{\rho} \over da} \; .
\ee
Using this expression to eliminate the scalar kinetic energy in the Friedmann equation and solving the resulting equation for $V$ gives
\be
\label{V1}
V = {1 \over 8 \pi G} \dot{H} +  {3 \over 8 \pi G} H^2 - \tilde{\rho} - {1 \over 6} a {d \tilde{\rho} \over da} \: .
\ee
Thus, we can explicitly determine the evolution of scalar field kinetic and potential energies from knowledge of the scale factor evolution together with the present matter and radiation densities.\footnote{Equations (\ref{sc1}) and (\ref{V1}) are also valid with multiple scalar fields, where the left side of (\ref{sc1}) is replaced by $\sum_i {1 \over 2} \phi_i^2$.}

\paragraph{Evolution in the phase of matter and dark energy domination}

For most of the history of the Universe, the cosmological expansion has (according to conventional models) been dominated by matter (including dark matter) and dark energy. Over this period, we have to a very good approximation that
\be
\tilde{\rho} = {\rho_{matter}^{now} \over a^3} = {3 H_0^2 \Omega_M \over 8 \pi G} {1 \over a^3}  \; ,
\ee
where we are taking the usual definition
\be
\Omega_M = \rho_{matter}^{now} {8 \pi G \over 3 H_0^2}
\ee
and $H_0$ is the present value of $\dot{a}/a$. 

In this case, our equations (\ref{sc1}) and (\ref{V1}) for the evolution of the scalar kinetic and potential energy become
\be
\label{sc2}
K(t) \equiv {1 \over 2} \dot{\phi}^2 = -{1 \over 8 \pi G} \dot{H} -{3 \over 16 \pi G} \Omega_M H_0^2 {1 \over a^3} 
\ee
and
\be
\label{V2}
V(t) = {1 \over 8 \pi G} \dot{H} +  {3 \over 8 \pi G} H^2 -{3 \over 16 \pi G} \Omega_M H_0^2 {1 \over a^3} \: .
\ee
Thus, the scalar field potential and kinetic energy evolutions are determined completely by $a(t)$ and $\Omega_M$ in the approximation that we can ignore radiation.

The positivity of the kinetic energy at the present time gives a constraint
\be
\Omega_M \le -{2 \over 3} {(\dot{H})_0 \over H_0^2} = {2 \over 3}(1 + q_0) \; ,
\ee
where 
\be
q_0 = - \left(\frac{\ddot{a} a}{\dot{a}^2}\right)_0
\ee
is the ``deceleration parameter''. The inequality becomes an equality if there is no scalar field kinetic energy, so for a given scale factor (e.g. deduced from observations) models with scalar fields will have a lower $\Omega_M$ than the standard $\Lambda$CDM model without scalars, assuming both fit the scale factor data. A measurement of the deceleration parameter thus provides a model-independent upper bound on the matter density (allowing arbitrary scalar fields and potentials). More generally, the positivity of scalar kinetic energy at general times implies that 
\be
\Omega_M \le \min\left(-{2 \over 3} {\dot{H} \over H_0^2}\right) \; ,
\ee
where the minimum is taken over the range of times where we have reliable scale factor data.

\paragraph{Deducing the potential}
\label{sec:single}

\begin{figure}
    \centering
    \includegraphics[scale = 0.8]{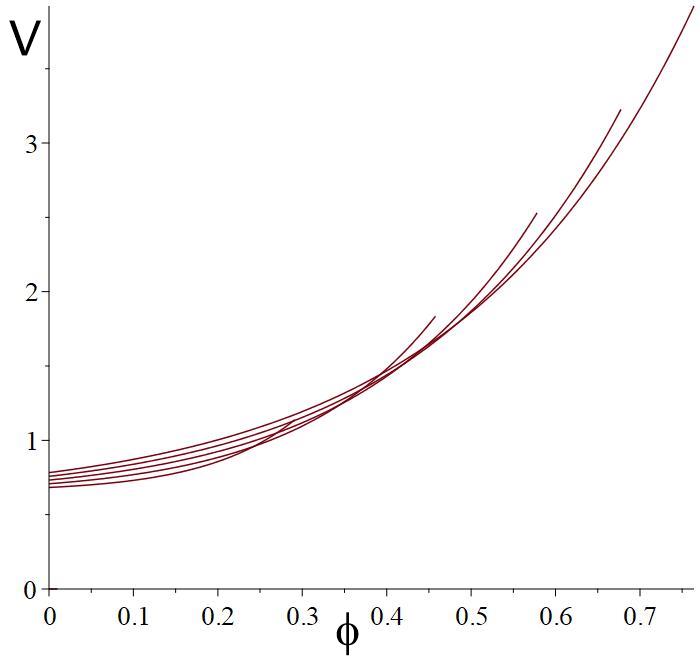}
    \caption{Potential $V(\phi)$ required to reproduce the $\Lambda$CDM fit to the observed scale factor back to $a=0.33$ ($z \approx 2$) in units with $8 \pi G/3 = 1$. We use the $\Lambda$CDM value $\Omega_M^0 = 0.334$ and plot the potentials required if $\Omega_M = 0.3,0.25,0.2,0.15$ and $0.1$ (bottom to top at the left). The scalar rolls from right to left, with the leftmost value representing the present time.}
    \label{fig:Vplots}
\end{figure}

Using the equation (\ref{sc2}) for the scalar kinetic energy, we can take the square root and integrate, to obtain
\be
\label{phisol}
\phi(t) = \phi(t_0)+\int_{t_0}^{t} d \hat{t} \sqrt{-{1 \over 4 \pi G} \dot{H}(\hat t) -{3 \over 8 \pi G} \Omega_M H_0^2 {1 \over a^3(\hat t)}} \: ,
\ee
where $t_0$ represents the present time, and we have chosen (without loss of generality) the convention that the scalar field is increasing its value at present. Since we also have
\be
\label{V3}
V(t) = {1 \over 8 \pi G} \dot{H} +  {3 \over 8 \pi G} H^2 -{3 \over 16 \pi G} \Omega_M H_0^2 {1 \over a^3} \; ,
\ee
we can combine (\ref{phisol}) and (\ref{V3}) to parametrically define $V(\phi)$. Thus, observations of $a(t)$ and $\Omega_M$ determine in principle the part of the scalar potential over which the scalar varies during the time for which $a(t)$ is known. If we know $a(t)$ but not $\Omega_M$, we have a one-parameter family of possible potentials that can reproduce this, where the parameter is the value of $\Omega_M$. As an example, scalar field evolution in the potentials shown in Figure \ref{fig:Vplots} with $\Omega_M = 0.3,0.25,0.2,0.15$ and $0.1$  gives precisely the same scale factor evolution as the best fit $\Lambda$CDM model with $\Omega_M = 0.334$. Details of this calculation are in Appendix A. This highlights that, to the extent the supernova data is well fit by a $\Lambda$CDM model, it is also well fit by other models with significant variation of $V$ during the recent cosmological history. On the other hand we will find that by allowing evolution of $V$ we can achieve a better fit to combine supernova and BAO data.

\paragraph{Rescaled variables}

In the transformation from $a(t)$ to $V(\phi)$, a rescaling of time translates to an overall rescaling of the potential. Thus, we can think of the Hubble parameter $H_{0}$ as setting the overall scale of the potential, and the remaining shape of $a(t)$ as determining the shape of $V$. To highlight this, it is convenient to define rescaled quantities (we will suppress the tildes below and always refer to these) 
\be
s = H_0 (t - t_0) \: , \qquad \tilde{H} = \frac{H}{H_{0}} \: , \qquad  \tilde{V}(\phi) = H_{0}^{-2} V(\phi) \: .
\label{eq:hubblerescaling}
\ee
If we also work in units where $8 \pi G/3 = 1$, the equations simplify to 
\be
\label{phisol1}
\phi(s) = \phi(0) + \int_0^{s} d s_1 \sqrt{-{2 \over 3} {d H \over ds} - {\Omega_M  \over a^3}}
\ee
and
\be
\label{V3a}
V(s) = {1 \over 3}  {d H \over ds} + H^2 - {\Omega_M  \over 2 a^3} \; ,
\ee
parametrically defining $V(\phi)$.

\section{Observational data for scale factor evolution} \label{sec:observational}

We have seen in the previous section that precise knowledge of the scale factor evolution $a(t)$ together with $\Omega_M$ would allow us to reconstruct the scalar potential $V(\phi)$ and the scalar field evolution $\phi(t)$. We would now like to understand how the scalar potential and evolution are constrained by existing observations.

\paragraph{Model-independent measurements of $\Omega_M$}
The mass density parameter $\Omega_M$ is rather weakly constrained if we don't assume a cosmological model. For example, recent attempts at  model-independent estimates are provided by the works \cite{Li:2019nux, Ruiz-Zapatero:2022zpx} that consider direct observations of the expansion and gravitational dynamics on the expanding background to place constraints on $\Omega_M$. The latter paper reports a range $\Omega_M = 0.224 \pm 0.066$. Allowing a 2 $\sigma$ range gives $\Omega_M \in (0.1,0.35)$, so it appears that the matter density parameter is only loosely constrained by direct observations. We will simply take $\Omega_M$ to be a free parameter in our models.

\paragraph{Measurements of $a(t)$}
The most direct observations of the scale factor evolution come from redshift vs brightness observations of type Ia supernovae. The state-of-the-art compendium of carefully studied SNe Ia is the so-called Pantheon+ sample (2021) \cite{Pan-STARRS1:2017jku, Scolnic:2021amr}, which draws on observations from a number of independent surveys (including CfA, Pan-STARRS, Sloan Digital Sky Survey, Supernova Legacy Survey, Carnegie Supernova Project, Hubble Space Telescope, etc.). These comprise 1701 ``spectroscopically confirmed" SNe Ia, between redshifts $z \approx 0$ and $z \approx 2.3$, with roughly a thousand SNe Ia at redshift $z > 0.1$. This data has been used to constrain cosmological parameters \cite{Brout:2022vxf} and the present Hubble constant \cite{Riess:2021jrx}. 

The supernova database provides results for observed magnitudes and redshifts of supernovae. Correcting for the effects of peculiar velocities (the velocities of objects relative to the cosmological background), the redshifts are related directly to the scale factor at the time of the supernova by
\begin{equation}
\label{ztoa}
    a = \frac{1} {1 + z} \: .
\end{equation}
The magnitudes are related fairly directly to luminosity distance $d_L$ (the distance that would be deduced from observed apparent magnitude and the presumed absolute magnitude assuming a time-independent flat geometry).\footnote{In the database, this information is presented as the distance modulus $\mu = 5 \log_{10}(d_L/10 \textnormal{pc})$.} This is algebraically related to the (positive) conformal time $\eta$ of the supernova (related to the usual cosmological time by $dt/a(t) = d \eta$) relative to the present time via
\begin{equation}
\label{dtoneta}
     \eta = d_{L} / (1 + z)  \: , 
\end{equation}
Thus, the relation between $1/(1+z)$ and $d_{L} / (1 + z)$ from supernova data can be considered as an observational picture of $a(\eta)$, the scale factor evolution in conformal time.

Another important probe of the scale factor evolution is the observation of baryon acoustic oscillations (BAOs). In this work, following the analysis in \cite{brout2022pantheon+}, we consider data from the Sloan Digital Sky Survey (SDSS) Main Galaxy Sample \cite{Ross:2014qpa}, the SDSS Baryon Oscillation Spectroscopic Survey (BOSS) \cite{BOSS:2016wmc}, the SDSS Extended Baryon Oscillation Spectroscopic Survey (eBOSS) Luminous Red Galaxies \cite{eBOSS:2020lta, eBOSS:2020hur}, SDSS eBOSS Emission Line Galaxies \cite{eBOSS:2020fvk}, SDSS eBOSS Quasars \cite{eBOSS:2020uxp, eBOSS:2020gbb}, and SDSS eBOSS Lyman-alpha forest \cite{eBOSS:2020tmo}. These surveys constitute the following data:
\begin{itemize}
        \item MGS: 63,163 galaxies with $0.07 < z < 0.2$ ($z_{\text{eff}} = 0.15$)
        \item BOSS: 
        1.2 million galaxies divided into three partially overlapping redshift slices within $0.2 < z < 0.75$ ($z_{\text{eff}} = 0.38, 0.51, 0.61$)
        \item eBOSS LRG: 377,458 galaxies with $0.6 < z < 1$ ($z_{\text{eff}} = 0.698$)
        \item eBOSS ELG: 173,736 galaxies with $0.6 < z < 1.1$ ($z_{\text{eff}} = 0.845$)
        \item eBOSS Quasars: 343,708 quasars with $0.8 < z < 2.2$ ($z_{\text{eff}} = 1.48$)
        \item eBOSS Ly$\alpha$: Auto-correlation of Ly$\alpha$, as well as cross-correlation with 341,468 quasars at redshifts $z > 1.77$ ($z_{\text{eff}} = 2.33$). 
\end{itemize}
The data is packaged as a geometric measure, often a comoving distance, at the effective redshift $z_{\text{eff}}$. 

BAO measurements alone are only sensitive to the dimensionless product $H_{0} r_{\text{d}}$, where $r_{\text{d}}$ is the sound horizon at the drag epoch, rather than to $H_{0}$. Consequently, factors of $r_{\text{d}}$ will appear in distance measurements obtained by the above BAO surveys. 
These distance measurements are specified either directly by the Hubble rate multiplied by the sound horizon, $r_{\text{d}} H(z_{\text{eff}})$, the comoving distance divided by the sound horizon,
\begin{equation}
        \frac{d_{M}(z_{\text{eff}})}{r_{\text{d}}} = \frac{c}{r_{\text{d}}} \eta(z_{\text{eff}})
\end{equation}
or the volume-averaged angular diameter distance divided by the sound horizon, which can be written as
\begin{equation}
    \frac{d_{V}(z_{\text{eff}})}{r_{\text{d}}} = \frac{1}{r_{\text{d}}} \left[ \frac{c z_{\text{eff}} (1+z_{\text{eff}})^{2}}{H(z_{\text{eff}})} d_{A}^{2}(z_{\text{eff}}) \right]^{1/3} = \frac{1}{r_{\text{d}}} \left[ \frac{c^{3} z_{\text{eff}}}{H(z_{\text{eff}})} \eta^{2}(z_{\text{eff}}) \right]^{1/3} \: .
\end{equation}
Some of these expressions involve the assumption that the cosmology is spatially flat.

\subsection{Constraining models based on supernova data} \label{sec:constrain_SN}

We would like to use the supernova data to constrain cosmological models with a single scalar field. For this, we follow the methodology of the Pantheon+ cosmology analysis.

The Pantheon+ supernova database provides values for redshift $z_i$ (corrected for known peculiar velocities) and a ``distance modulus'' $\mu_i$ deduced from the supernova light curve and related to the inferred luminosity distance by 
\be
\label{defmu}
\mu = 5 \log_{10} (d_L / 10\textnormal{pc}) \; .
\ee

For a choice of parameters $H_0$, $\Omega_M$, $V_i$, and $M$ we can obtain a model value $\mu_i$ for the redshift $z_i$ corresponding to each supernova by assigning a scale factor
\be
a_i = {1 \over 1 + z_i}  
\ee
and numerically solving the coupled scalar evolution equations and Friedmann equations to determine $\eta(a)$ where we define $\eta$ to be the (positive) conformal time into the past in units of $1/H_0$. The relevant equations for the rescaled variables can be written in first order form as
\be
\label{ODEs}
{d \phi \over d a} = - {v_\phi \over a H}  \qquad
{d v_\phi \over da} = {V'(\phi) \over a H} - 3 {v_\phi \over a}  \qquad 
{d \eta \over da} =  - {1 \over a^2 H} \qquad H = \sqrt{{\Omega_M \over a^3} + {1 \over 2} v_\phi^2 + V(\phi)}
\ee
with boundary conditions
\be
\eta(1) = 0 \qquad \phi(1) = 0 \qquad v_\phi(1) = \sqrt{2(1 - \Omega_M - V(0))} \; .
\ee
The model value $\mu_i$ is then computed from (\ref{defmu}), where using (\ref{dtoneta})
\be
(d_L)_i = {\eta(a_i) \over a_i H_0}   \; .
\ee
To evaluate the fit of a certain set of model parameters, we calculate a $\chi^2$ value as
\be
\label{chisquared}
\chi^2 = (\mu_{data} - \mu_{model})_i C^{-1}_{ij} (\mu_{data} - \mu_{model})_j 
\ee
where $C$ is the covariance matrix provided by the Pantheon+ results that takes into account statistical and systematic error. Here 
\be
(\mu_{data})_i = (m_B)_i - M
\ee
where $m_B$ is the measured (and corrected) apparent magnitude (see \cite{brout2022pantheon+} for a detailed explanation) and M is the SnIa absolute magnitude, taken as a free parameter.

According to the standard analysis, the $\chi^2$ is used to assign a likelihood
\be
{\cal L} = e^{- \chi^2 /2}
\ee
to a set of model parameters. We start from an assumed prior distribution of parameters which we take to be flat apart from the constraints $\Omega_M > 0$ and $V_0 + \Omega_M < 1$, the latter required by the last equation in \ref{ODEs} which is the constraint of non-negative kinetic energy for the scalar field. In our analysis, we use a Markov Chain Monte Carlo (MCMC) algorithm to sample from the resulting probability distribution, yielding histograms and correlation plots for the various parameters.

As described in the Pantheon+ cosmology analysis, the Hubble parameter $H_0$ is degenerate with the absolute magnitude parameter $M$. In order to constrain $H_0$, the Pantheon+ analysis also considered a modified likelihood function that takes into account additional data from the SH0ES survey \cite{riess2022comprehensive} for the distances to certain host galaxies deduced from using Cepheid variable stars as a standard candle. The modified $\chi^2$ is defined by replacing $(\mu_{data} - \mu_{model})_i$ in equation (\ref{chisquared})  with $(\mu_{data} -\mu_{Cepheid})_i$ for supernovae in galaxies with an independent distance estimate. In this case, the fit constrains $M$ and $H_0$ separately since it takes into account both how well the model matches with the supernova observations but also how well the supernova data match with the independent distance measure from the Cepheids.

We mention finally that the Pantheon+ analysis found that the best fit $\Lambda$CDM models have residuals $\mu - \mu_{model}$ that are significantly positive on average for very nearby ($z < 0.01$) supernovae. This is taken to indicate the likely presence of some local effect (e.g. unmeasured peculiar velocities due to a local density excess) that has not been accounted for.
To ensure that these local effects are not having a significant effect on the results for our model parameters, we typically employ a $z<0.01$ cut apart from data with Cepheid calibrators. We have compared this with an alternative treatment of low $z$ data
considered in \cite{Perivolaropoulos:2023iqj} which uses all data but introduces an independent  absolute magnitude parameter for SNe Ia with $z$ smaller than some $z_{crit}$. This gives similar results for the main  parameters of interest, as we describe below.

As a check on our procedure, we performed an analysis of the $w$CDM model, finding distributions for parameters in excellent agreement with the Pantheon+ results in \cite{brout2022pantheon+}. 

\subsection{Constraining models with BAO data}

For part of the analysis, we would like to combine the supernova analysis described in the previous subsection with additional constraints coming from the BAO data. As mentioned before, the BAO data sets provide measurements of distance at an effective redshift. These will be in the form of a measurement of $r_{\text{d}} H(z_{\text{eff}})$, $\frac{d_{M}(z_{\text{eff}})}{r_{\text{d}}}$, or $\frac{d_{V}(z_{\text{eff}})}{r_{\text{d}}}$. 

We can obtain a model value for the appropriate quantity as in Section \ref{sec:constrain_SN}, by solving a system of coupled ODEs, though now for a given scale factor $a_{\text{eff}} = \frac{1}{1+z_{\text{eff}}}$ we may be computing the associated Hubble rate, comoving distance, or volume-averaged distance rather than the luminosity distance. These quantities can be computed given $\eta(a)$ and $H(a)$ for the model. 

While the sound horizon at recombination, which is closely related to $r_{\text{d}}$, has been very precisely inferred from the angular scale of acoustic peaks in the cosmic microwave background as observed by Planck \cite{Planck:2018vyg}, this determination assumes a $\Lambda$CDM cosmology. 
Since we do not assume that the linear and quadratic scalar potential models considered here, let alone the $\Lambda$CDM model, remain valid during the entire expansion history since recombination, we will float the value of $r_{\text{d}}$ as a global parameter in the following analysis. 

As in \cite{brout2022pantheon+}, we use the BAO likelihoods available in the CosmoSIS framework \cite{Zuntz:2014csq}, adapted to the class of scalar dark energy models considered here. 
Many, though not all, are Gaussian likelihoods; see the relevant analyses \cite{Ross:2014qpa, BOSS:2016wmc, eBOSS:2020lta, eBOSS:2020hur, eBOSS:2020fvk, eBOSS:2020uxp, eBOSS:2020gbb, eBOSS:2020tmo} and documentation in CosmoSIS for further details.

\section{Results} \label{sec:results}

In this section, we present our results for the parameter space of scalar potentials giving rise to scale factor evolution in good agreement with observations. 

\subsection{Linear potentials: supernovae only} \label{sec:results_linear}

We consider first a model with linear potential $V(\phi) = V_0 + V_1 \phi$. Even for a more complicated potential, expanding about the present value of $\phi$ will give this linear model for a short enough range of $\phi$, corresponding to some sufficiently small window of cosmological time. It is possible that this applies to the window $z < 2$, but this is not necessarily the case; we extend the analysis to include quadratic terms in the potential in the next section. One motivation for studying the linear potential model is that it will give us an idea of how far we can deviate from a $\Lambda$CDM model while still obtaining a good fit to the data. In particular, it will provide a lower bound on the acceptable amount of variation of the potential energy during the evolution. 

\begin{figure}
    \centering
    \includegraphics[width=17cm]{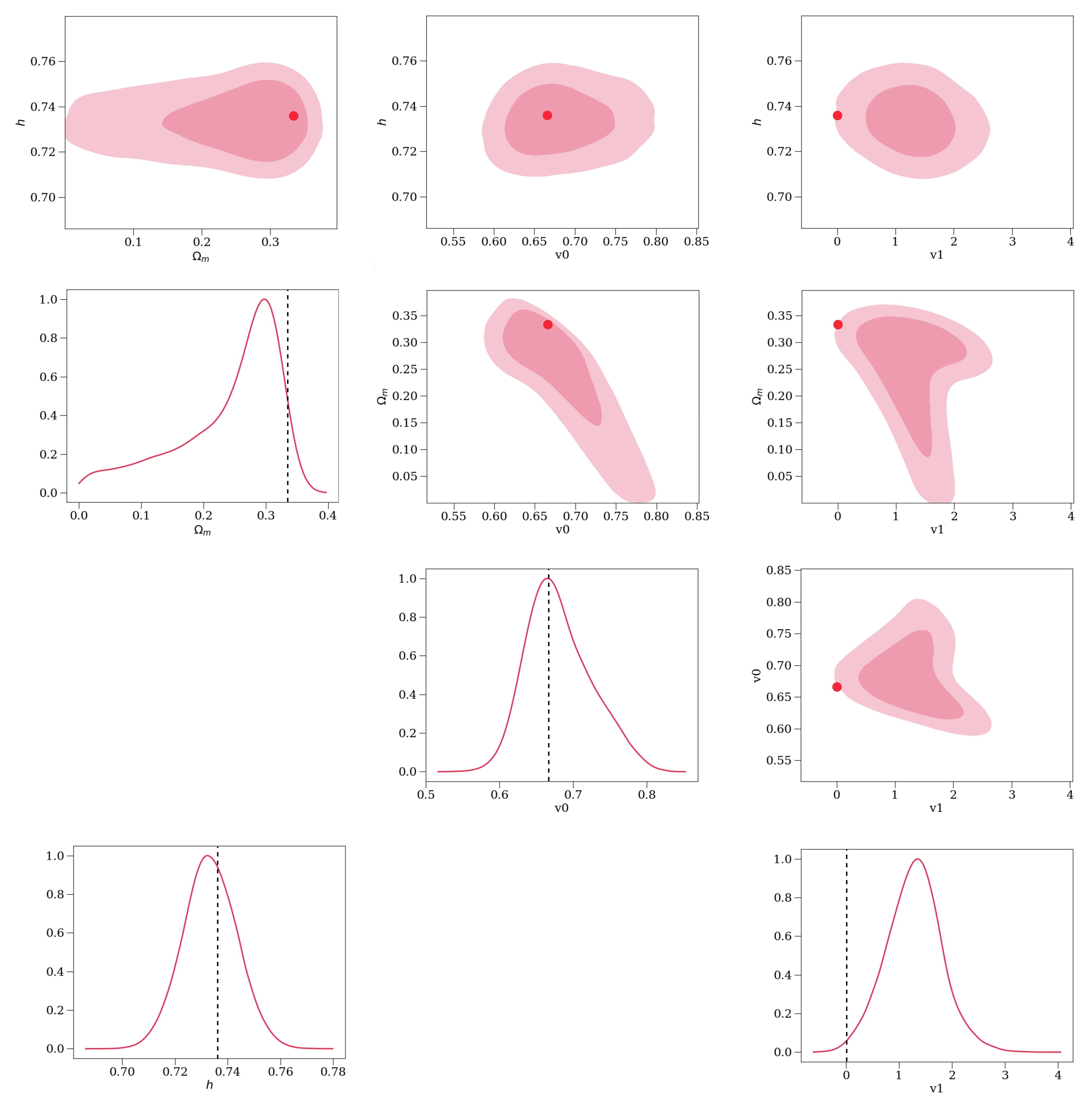}
    \caption{Distribution of parameter values $h$, $\Omega_M$, $V_0$, and $V_1$ for a linear potential scalar field model, using $e^{-\chi^2 / 2}$ likelihood calculated from Pantheon+ data calibrated with SH0ES Cepheid distances (68\% and 95\% contours shown). We define $h$ by $H_{0} = 100 h \: \text{km} \: \text{s}^{-1} \text{Mpc}^{-1}$. The $\Lambda$CDM model corresponds to $V_0 = 1 - \Omega_M$, $V_1 = 0$; the $\Lambda$CDM parameter values found by Pantheon+ \cite{brout2022pantheon+} are demarcated by a red dot or a dashed black line in the figure.}
    \label{fig:V01}
\end{figure}

In Figure \ref{fig:V01}, we show histograms and distributions for the parameters $\Omega_M$, $H_0$, $V_0$ and $V_1$, making use of the Cepheid data and the $z < 0.01$ cut as in the Pantheon+ analysis. We are using the rescaled variables in which the potential is normalized as $V \to V/H_0^2$ so that $V_0 = \Omega_\Lambda$ in a $\Lambda$CDM model. We are taking units with $8 \pi G / 3 = 1$ so $V_1$ of order 1 means that the rescaled potential changes by an order one amount when $\phi$ changes by a Planck scale amount.

\begin{figure}
    \centering
    \includegraphics[width=\textwidth]{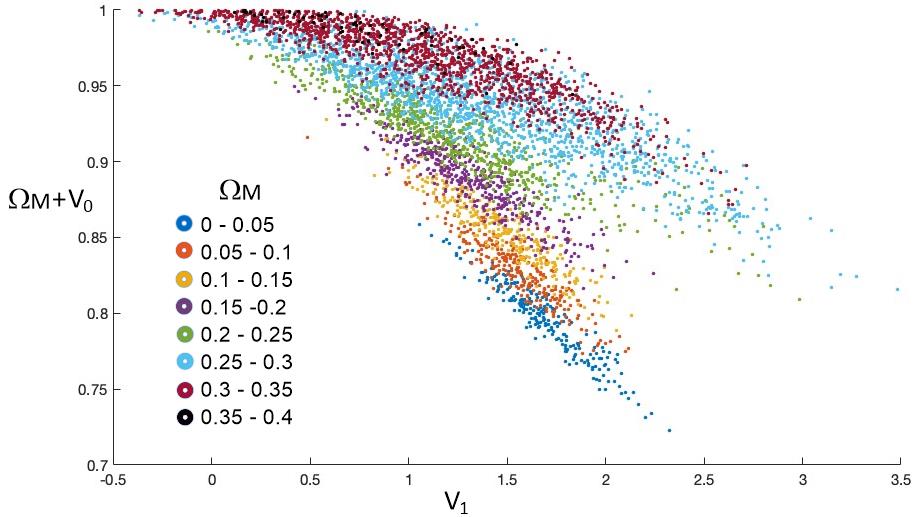}
    \caption{ $\Omega_M + V_0$ vs $V_1$ for a linear potential scalar field model, using $e^{-\chi^2 / 2}$ likelihood calculated from Pantheon+ data calibrated with SH0ES Cepheid distances. The $\Lambda$CDM model corresponds to $V_0 + \Omega_M = 1$, $V_1 = 0$.}
    \label{fig:OMplusV0vsV1}
\end{figure}
In Figure \ref{fig:OMplusV0vsV1}, we show $\Omega_M + V_0$ vs $V_1$, colour coded to show the $\Omega_M$ values. We see that the $\Lambda$CDM model, with $\Omega_M + V_0 = 1$ and $V_1 = 0$ is on the tail of the distribution defined by the $\exp(-\chi^2/2)$ likelihood. Models in the distribution typically have a significant variation in the scalar potential energy over the time range corresponding to the data: on average, we find $\textnormal{Range}[V(\phi)]/V_0 \approx  0.97$ so there is typically an ${\cal O}(1)$ variation in the potential during the evolution for models that provide a good fit to the data. For linear potential models, the potential will eventually descend to negative values. Based on the distribution produced by our MCMC sampling, we find that the medial time for this to happen is approximately 1.1 Hubble times, a little larger than the current age of the universe.

From Figure \ref{fig:OMplusV0vsV1} we see that there appear to be two different branches of allowable model parameters at larger $V_1$. The upper branch reaching the largest values of $V_1$ corresponds to models where $\phi$ first moves upward on the potential and then falls back down. In the lower branch (corresponding to small $\Omega_M$ values), the evolution of $\phi$ is monotonic, with $V$ monotonically decreasing.  

We find that the marginalized distribution for $V_1$ has average value $V_1 \approx 1.3$ with the range $V_1 \in [0.72,1.77]$ containing 68\% of the distribution with an equal amount above and below. We find 99.1\% of the distribution has $V_1 > 0$, so the $\Lambda$CDM model is somewhat disfavoured within this space of linear potential models. This becomes more pronounced if we take into account the full data set (without the $z < 0.01$ cut). Here, we find a significantly larger average value $V_1 \approx 2.0$ with 68\% of the distribution in $[1.51,2.56]$ and $V_1 > 0$ for 99.7\% of the distribution. The preference for non-zero $V_1$ is already visible in results \cite{wang2004current, Sahlen:2006dn, Huterer:2006mv} based on older data sets, though the result is not nearly as significant.

For the present value of the rescaled potential (equal to $\Omega_\Lambda$ in a $\Lambda$CDM model), we find $V_0 = 0.681 \pm 0.044$.

For the $\Omega_M$ parameter, we find 99\% of the distribution lies in the range $\Omega_M \in [0,0.352]$. While the distribution is peaked near $\Omega_M = 0.3$, similar to the $\Lambda$CDM 
value of $0.334$ using the same data, the distribution is very broad (see Figure \ref{fig:V01}) extending to $\Omega_M = 0$ with significant weight.

Finally, for the Hubble parameter, we find $H_0 = (73.3 \pm 1.0)$ km/s/Mpc, in agreement with the Pantheon+ cosmology analysis \cite{brout2022pantheon+}. 

\begin{table}
\begin{center}
\begin{tabular}{|| c | c | c | c ||}
 \hline
 $h$ & $\Omega_{M}$ & $V_{0}$ & $V_{1}$ \\ [0.5ex] 
 \hline\hline
 $0.733 \pm 0.010$ & $0.24^{+0.10}_{-0.03}$ & $0.681^{+0.034}_{-0.053}$ & $1.3 \pm 0.5$ \\
 \hline
\end{tabular}
\caption{Summary of marginalized parameter constraints for the linear potential model, using supernova data only with a $z < 0.01$ cut.  }
\label{tab:params}
\end{center} 
\end{table}

A summary of the marginalized parameter values and their uncertainties, assuming the $z<0.01$ cut, can be found in Table \ref{tab:params}. As we described above, as a check on the sensitivity to local effects we have also followed \cite{Perivolaropoulos:2023iqj} in fitting distinct parameters $M_{1}, M_{2}$ for the absolute magnitude of SNe Ia with $z$ smaller than or larger than $0.005$ respectively. We find very comparable results
\begin{equation}
    h = 0.732 \pm 0.011 \: , \quad \Omega_{M} = 0.25_{-0.03}^{+0.09} \: , \quad V_{0} = 0.676_{-0.049}^{+0.031} \: , \quad V_{1} = 1.3_{-0.4}^{+0.5} \: .
\end{equation}




\subsection{Linear potentials: supernovae + BAOs} \label{sec:results_linear_bao}

\begin{figure}
    \centering
    \includegraphics[width=15cm]{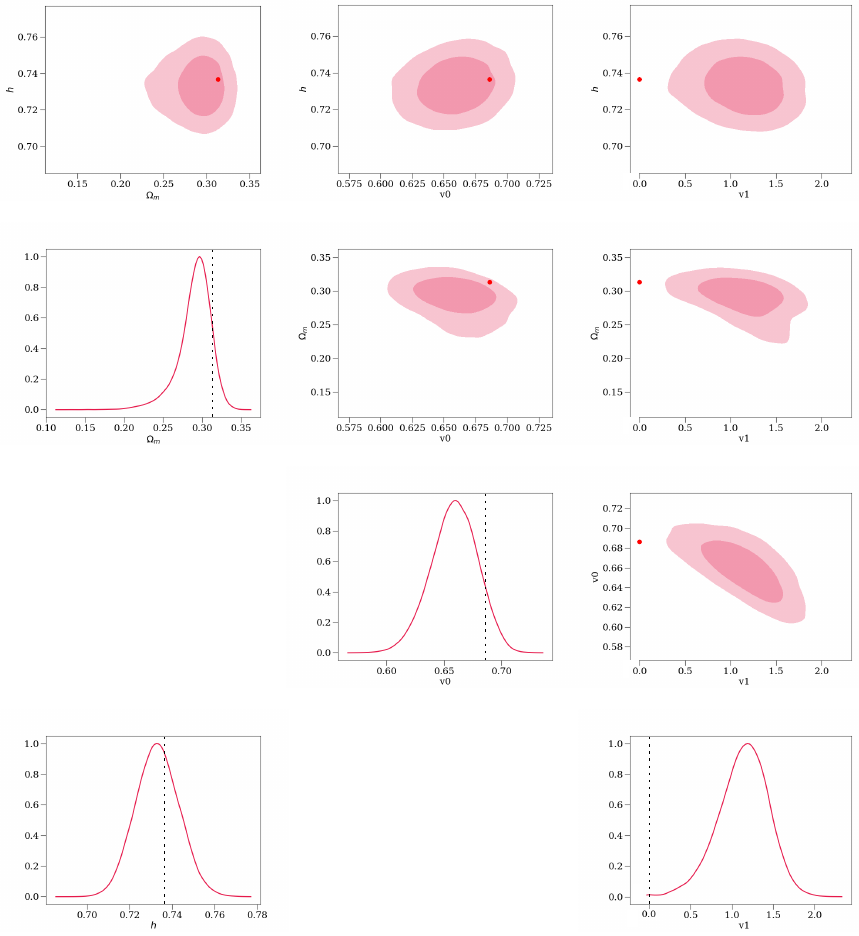}
    \caption{ Distribution of parameter values $h$, $\Omega_M$, $V_0$, and $V_1$ for a linear potential scalar field model, using Pantheon+ data calibrated with SH0ES Cepheid distances and BAO data (68\% and 95\% contours shown). We define $h$ by $H_{0} = 100 h \: \text{km} \: \text{s}^{-1} \text{Mpc}^{-1}$. The $\Lambda$CDM model corresponds to $V_0 = 1 - \Omega_M$, $V_1 = 0$; the best fit $\Lambda$CDM parameter values are demarcated by a red dot or a dashed black line in the figure.}
    \label{fig:LinearPotential_Results_SNBAO}
\end{figure}

We again consider the linear potential model of the previous subsection, now incoporating BAO data. We apply the $z < 0.01$ cut on supernova data throughout this section. As in the previous subsection, we use rescaled variables and set $8 \pi G/3=1$. The results are shown in Figure \ref{fig:LinearPotential_Results_SNBAO}. 

The average value of the marginalized distribution for $V_{1}$ is now the slightly lower value $V_{1} \approx 1.13$, with the smaller range $V_{1} \in [0.88, 1.46]$ containing 68\% of the distribution. We now find that approximately $99.99\%$ of the entire distribution of sampled values is supported on $V_{1} > 0$, implying that $\Lambda$CDM is significantly disfavoured within this space of models.\footnote{In order to establish this fraction, we performed a MCMC run restricting $V_1$ to values below 0.3 with enough data that $~1000$ points had $V_1 < 0$. We then estimated the fraction of points in the full distribution with $V_1 < 0$ as $p(V_1 < 0) = p(V_1 < 0 | V_1 < 0.3) p(V_1 < 0.3)$ where $p(V_1 < 0.3)$ was estimated from an unconstrained simulation.}

While the $\exp(-\chi^2/2)$ distribution has very little weight for $V_1 < 0$, we note that for particular choices of parameters in the $\Lambda$CDM model, it is possible to obtain $\chi^2$ values that are only greater than than those with the best fit linear potential model by an amount $\Delta \chi^2 \approx 3.7$. For a Gaussian distribution, this $\Delta \chi^2$ would be much larger for points so far out on the tail of the distribution. This indicates that there is only a very small volume of parameter space near the best fit $\Lambda$CDM models that have a comparable $\Delta \chi^2$ to the best fit model, while a much larger volume of parameter space near the best fit linear potential model has a comparable $\Delta \chi^2$ to that model.\footnote{This may be at least partly explained by the observation that $\Omega_M$ and $V_0$ appear to be strongly correlated for small $V_1$ for models with low $\chi^2$.} Because of this feature, alternative model selection criteria based only on the minimum $\chi^2$ values do not significantly favour the linear potential models over $\Lambda$CDM. For example, according to the Akaike Information Criterion the preferred model has the larger value of  $AIC = 2k - 2 \ln ({\cal L})$ where $k$ is equal to the number of parameters. Since the linear potential model has two more parameters than $\Lambda$CDM, we find  $AIC_{\Lambda CDM} - AIC_{linear} \approx 0.3$. This indicates a very slight preference for $\Lambda$CDM according to the criterion. A related model selection test, the Baysian Information Criterion, replaces $AIC$ with $BIC = k \ln(N) - 2 \ln ({\cal L})$ where $N$ is the number of samples ($\sim 1500$) in our case. This penalizes additional parameters more heavily. Here, we find $BIC_{\Lambda CDM} - BIC_{linear} \approx 8$ and the criterion would give preference to the $\Lambda$CDM model. In summary, while the standard likelihood analysis appears to significantly prefer models with non-trivial scalar evolution, criteria that do not take into account parameter space volumes still are more neutral or prefer $\Lambda$CDM. This ambiguity is likely to be resolved by comparison with improved data sets that will be available in the relatively near future.

We find that the typical range of scalar evolution averaged over the distribution is $\text{Range}[V(\phi)]/V_{0} \approx 0.36$; this is notably smaller than in the analysis without BAO, though still $O(1)$. The time taken for the scalar field to descend to negative values of the potential is consequently somewhat longer, now approximately $2.1$ Hubble times. 

The current value of the rescaled potential $V_{0}$ is found to be $V_{0} = 0.659 \pm 0.020$, comparable to the value found without BAO and to the value of $\Omega_{\Lambda}$ deduced for the flat $\Lambda$CDM model by Pantheon+ \cite{brout2022pantheon+}. The current value of the Hubble parameter is found to be $H_{0} = (73.3 \pm 1.0) \: \text{km/s/Mpc}$, identical to the value found without BAO and consistent with the Pantheon+ analysis. 

The peak value of the marginalized distribution of the matter density is again near $\Omega_{M} = 0.3$; however, the distribution is truncated substantially at small values compared to the analysis without BAO, with 68\% confidence interval now given by $\Omega_{M} \in [0.278, 0.313]$. 

A summary of the marginalized parameter values and their uncertainties can be found in Table \ref{tab:params_BAO}. 

\begin{table}
\begin{center}
\begin{tabular}{|| c | c | c | c ||}
 \hline
 $h$ & $\Omega_{M}$ & $V_{0}$ & $V_{1}$ \\ [0.5ex] 
 \hline\hline
 $0.733 \pm 0.010$ & $0.291^{+0.023}_{-0.013}$ & $0.659^{+0.021}_{-0.019}$ & $1.1 \pm 0.3$ \\
 \hline
\end{tabular}
\caption{Summary of marginalized parameter constraints for the linear potential model, based on supernova and BAO data. More detailed information about the distribution of parameter values can be found in the main text of Section \ref{sec:results_linear_bao} and in Figure \ref{fig:LinearPotential_Results_SNBAO}. }
\label{tab:params_BAO}
\end{center} 
\end{table}

\subsection{Quadratic potentials} \label{sec:results_quadratic}

Next, we consider adding a quadratic term to the potential, taking $V(\phi) = V_0 + V_1 \phi + V_2 \phi^2 /2$. We can view this as including the next term in the Taylor series approximation to some more general potential about the present value of the scalar field. More generally, it will provide insight into the model behaviors that are possible with potentials that have a changing slope. In this section, which is intended to be an exploration of the parameter space rather than a full analysis of the constraints on parameters, we focus on the implications of supernova data.

For the space of models with parameters $\Omega_M, H_0, V_0, V_1, V_2$ and the absolute brightness adjustment parameter $M$, we find that it is possible to obtain a good fit to the supernova data (comparable to or better than the best fit $\Lambda$CDM model) for a large variety of potential parameters. In fact, the Markov Chain Monte Carlo analysis does not converge well since $V_2$ does not have a bounded range in models that provide a good fit to the data. A partial explanation is that starting from a $\Lambda$CDM model that provides a good fit to the data, we can set $V_2$ to whatever we want without any change in the evolution provided that the scalar field continues to sit at the extremum of the potential. However, we can also have models with non-trivial scalar evolution giving significant variation in the potential for essentially any $V_2$. 

For large positive values of  $V_2$, the scalar field can oscillate many times about the minimum of the potential. As we review in Appendix  \ref{sec:oscillation} below, such an oscillating scalar contributes to the energy density in the same way as non-relativistic matter plus a cosmological constant, up to corrections suppressed by powers of $H_0/\sqrt{V_2}$ \cite{turner1983coherent}. For large negative values of $V_2$, we can have a situation where the scalar spends a significant fraction of the time near the maximum of the potential, but changes significantly for early or late times.

\begin{figure}
    \centering
    \includegraphics[width= 0.85 \textwidth]{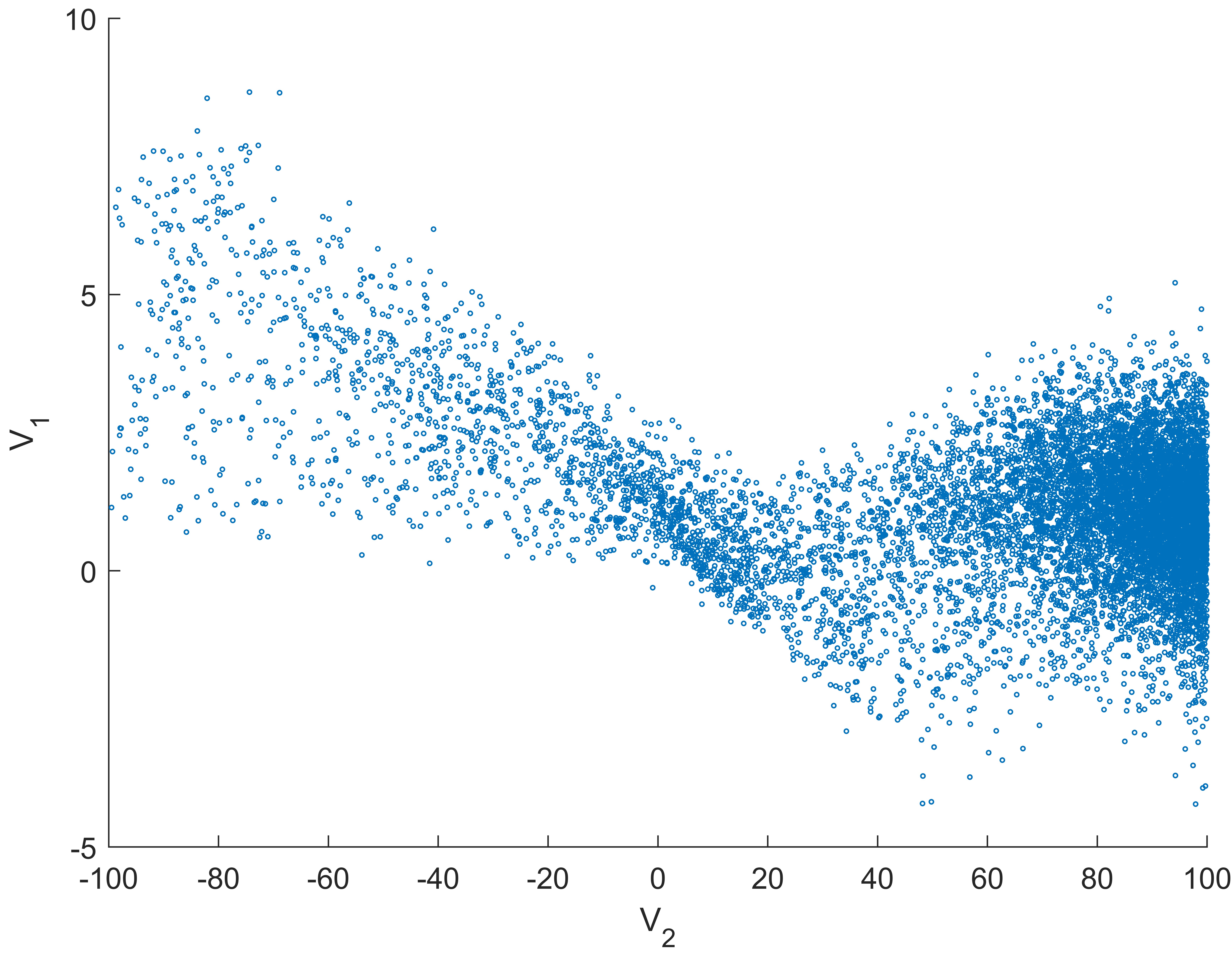}
    \caption{The ${\rm exp}(-\chi^2/2)$ distribution of potential parameters $V_1$ and $V_2$ for quadratic scalar potential model with restriction $|V_2|< 100$. The large accumulation of points for larger $V_2$ is the result of a degeneracy where some amount of matter is replaced by scalar field oscillating in the quadratic potential.}
    \label{fig:V012}
\end{figure}

\begin{figure}
    \centering
    \includegraphics[width= 0.7 \textwidth]{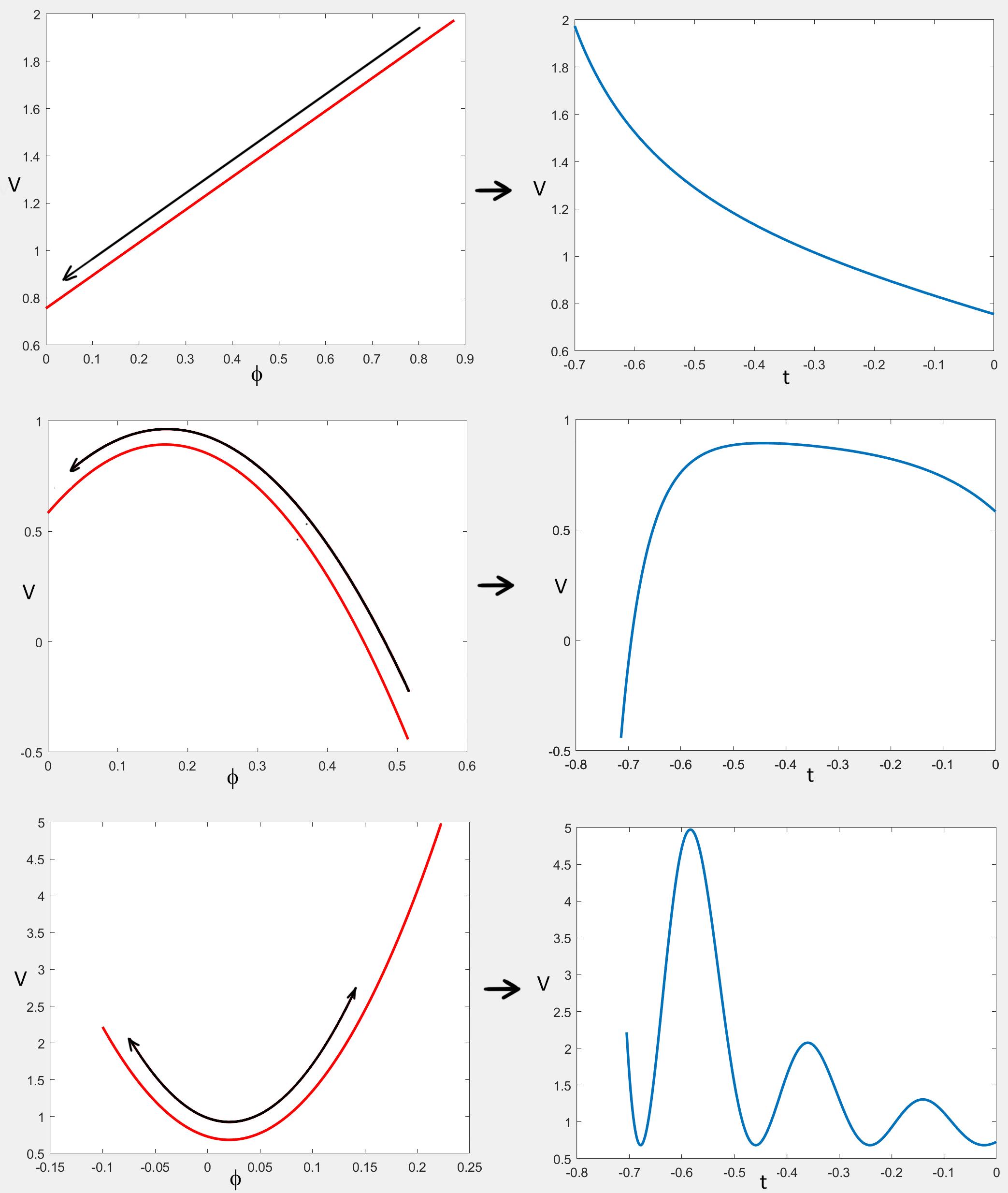}
    \caption{Potentials $V(\phi)$ with corresponding scalar potential energy evolution $V(t)$ for various qualitatively different models providing a good fit to the supernova data. Time is measured in units of $1/H_0$.}
    \label{fig:Vexamples}
\end{figure}

\begin{figure}
    \centering
    \includegraphics[width= 1.0 \textwidth]{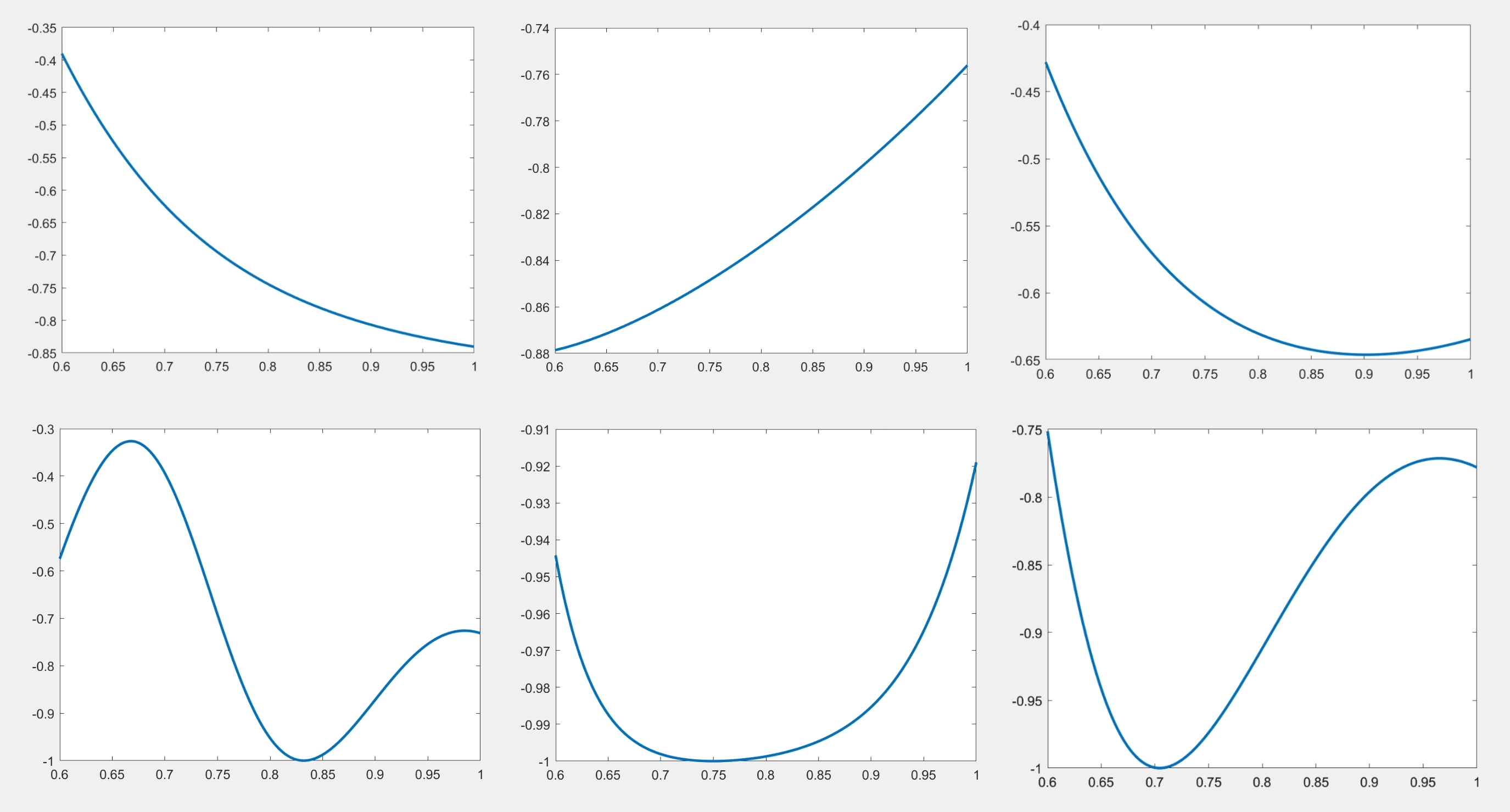}
    \caption{Equation of state parameter $w$ vs scale factor $a \in [0.6,1]$ for various models with a linear potential (top row) and a quadratic potential (bottom row), each providing a good fit to the supernova data.}
    \label{fig:wvsa}
\end{figure}

The results of an MCMC analysis with the restriction $|V_2| < 100$,  are shown in Figure \ref{fig:V012}. The accumulation of points at larger $V_2$ values reflects the degeneracy of replacing matter with oscillating scalar. In Figure \ref{fig:Vexamples}, we provide examples of various potentials and scalar evolutions that provide a fit to the data that is as good or better than $\Lambda$CDM. In Figure \ref{fig:wvsa}, we show the ``equation of state parameter'' $w = (\dot{\phi}^2/2 - V)/(\dot{\phi}^2/2 + V)$ as a function of the scale factor for several examples of models in our distributions for the linear potential and quadratic potential cases. We see a significant variety of behaviors here, so it is clear that the class of models we consider don't map neatly onto the $w$CDM models (where $w(a) = w_0$) or $w_{0}w_{a}$CDM models (where $w(a) = w_{0} + w_{a}/a$) considered in \cite{brout2022pantheon+} and various other analyses.

Allowing larger values of $|V_2|$, we can find examples with even lower $\chi^2$ than the examples with the restriction $|V_2| < 100$. A curious observation is that the example with the smallest $\chi^2$ we have found,
\be
\label{example}
\Omega_M = 0.339 \qquad H_0 = 71.1 \: \textnormal{km/s/Mpc} \qquad V_0 = 0.000845 \qquad V_1 = 29.1 \qquad V_2 = -562.1
\ee
(with $\Delta \chi^2 \approx 17$ compared to $\Lambda$CDM), the potential is a downward parabola, and the scalar remains near the maximum of the downward parabola until relatively late times ($\sim$1 Gyr before present) before descending to a value of nearly $V=0$ at present (the potential becomes negative within ${\cal O}(10^5)$ years). In this model, the universe actually stopped accelerating 0.4 Gyr ago! That the present value of the potential is so nearly $V=0$ would seem to be purely a coincidence.

If we impose a cut on the data of $z < 0.01$, the advantages of this or similar models over $\Lambda$CDM become less significant. A plausible explanation is that there is some unaccounted for local effect that makes the very nearby supernovae less reliable for deducing the recent scale factor evolution. In this case, the data would be expected to deviate from the correct model for background evolution, and the better fit for the quadratic potential model would simply be due to having extra parameters for the background evolution that can reproduce this local effect. On the other hand, if the $z < 0.01$ data are providing a good picture of the recent scale factor evolution, our observations with the quadratic potential model suggest that the discrepancy with the $\Lambda$CDM model for these recent data might be due to a recent decrease in the potential energy.

\begin{figure}
    \centering
    \includegraphics[width= 0.7 \textwidth]{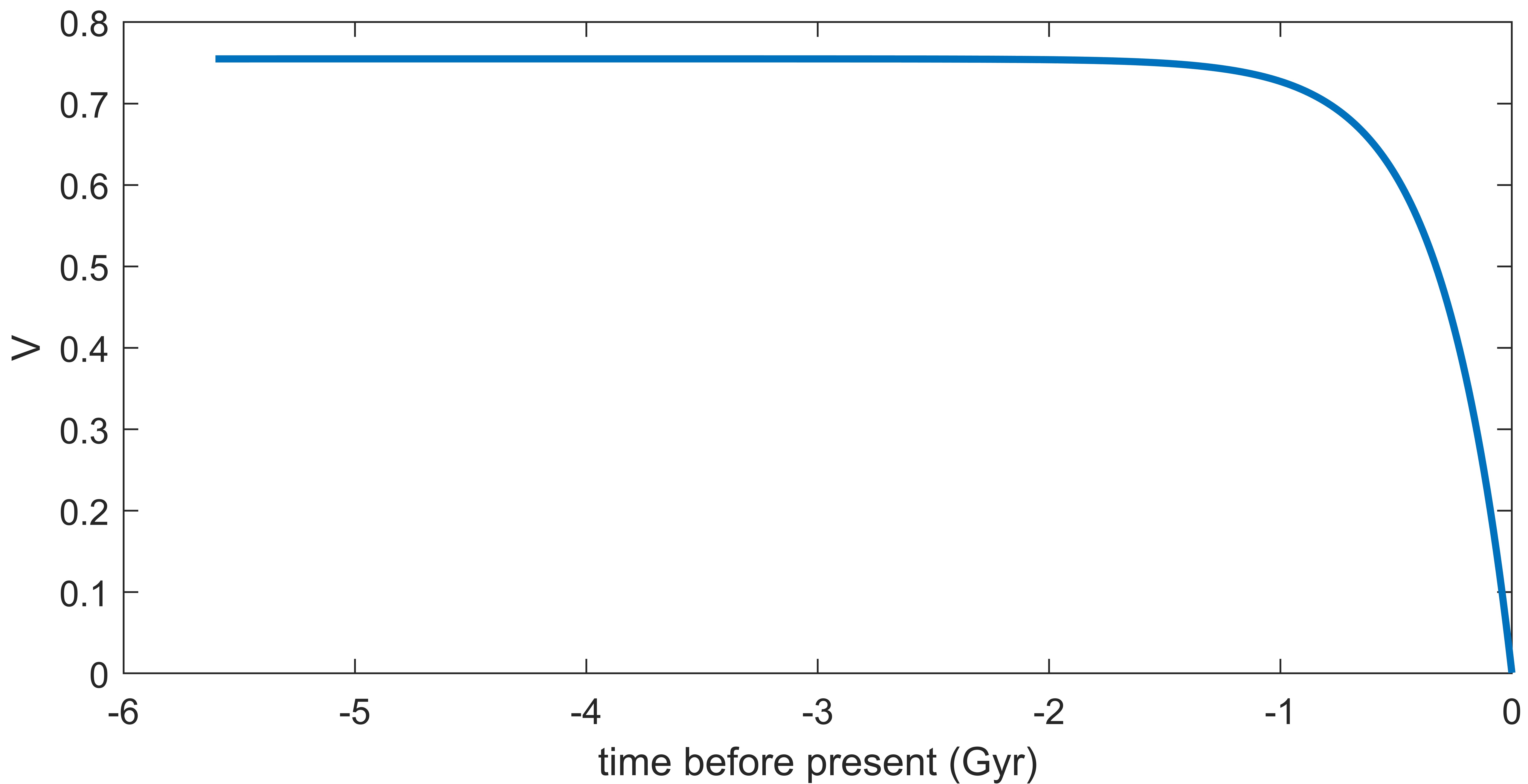}
    \caption{Scalar potential vs time for quadratic potential model (\ref{example}) providing the best fit to the Pantheon+ supernova data. The scalar field remains near the maximum of the potential until relatively recently.}
    \label{fig:PotentialVariation}
\end{figure}

\section{Discussion} \label{sec:discussion}

In this paper, motivated by theoretical considerations \cite{VanRaamsdonk:2022rts} suggesting that non-trivial scalar field evolution on cosmological time scales is natural, we have explored the extent to which our most direct observations of scale factor evolution are compatible with a varying dark energy arising from a time-dependent scalar field. 

One main conclusion is that there is a lot of room for such models, and models that are consistent with observations are not necessarily close to $\Lambda$CDM. Within the space of models we consider, we have found that in models providing a good fit to the data, the scalar potential typically changes by an order one amount compared to its present value during the time scale corresponding to the supernova data, roughly half the age of the universe. In the context of models with a linear potential, nearly the entirety of the distribution defined by the $\exp(- \chi^2/2)$ likelihood have a potential that is presently decreasing with time. One might have expected that if $\Lambda$CDM is actually the correct model, extending the parameter space to the linear potential models would yield a distribution with a similar fraction of models with $V_1 < 0$ and $V_1 > 0$. The dominance of $V_1 > 0$ models in our distribution could thus suggest that the correct model is not $\Lambda$CDM but one with a decreasing dark energy.

We should emphasize that there is no particular reason to believe that a linear approximation to the scalar potential should be valid for the entire time span $z < 2$ that we focus on. However, if the universe does indeed have some significant scalar evolution in the recent past, approximating the potential near the present value as a line with a general slope would be expected to do significantly better than approximating the potential by a constant. The fact that the likelihood analysis finds a significant preference for $V_1 > 0$ in the linear potential model thus suggests that the correct picture may be some model with significant recent scalar evolution with decreasing potential energy. On the other hand, our results should be interpreted with caution, since as we have explained, the $\Delta \chi^2$ between $\Lambda$CDM and the linear potential model is not as large as would be expected for a Gaussian distribution given the likelihood results. At the least, we hope the analysis presented here strongly motivates careful analyses of how the linear potential approximation models compare with the improved data in present and near-future supernovae and BAO surveys. 

For models with a quadratic potential, we find a variety of qualitatively different possibilities, including models where the scalar field is now descending a downward-pointing parabola and models where the scalar is oscillating in an upward-facing parabola. Again, the best fit model has a decreasing dark energy at late times. The simple model we consider could be extended in various ways, for example by considering multiple scalars possibly with a non-trivial metric on target space, or by considering non-minimal couplings to gravity.

Models with scalar fields varying on cosmological time scales introduce various other phenomenological constraints that must be satisfied. The quanta of such a field will be nearly massless and thus result in long range forces for any matter that they couple to. In order to avoid observed violations of the equivalence principle, couplings to ordinary matter should be extremely small, so there should be some theoretical explanation for why this would be the case. Such a small coupling may be more natural with a pseudoscalar axion-like field \cite{marsh2016axion}.

\subsection{Relation to $w_0w_a$ models and DESI results}

After the original version of this paper, the analysis \cite{DESI:2024mwx} appeared which also presents evidence for evolving dark energy based on supernova and BAO data. In that paper, it is shown that $\Lambda$CDM is disfavoured in the context of a $w_0w_a$ model for dark energy evolution, where the $w$ parameter is assumed to vary with scale factor in a linear way: $w(a) = w_{0} + w_{a}(1-a)$. 

We would like to emphasize that this is distinct from the linear potential models we consider here. In particular, the $w_{0}w_{a}$ models are not based on any underlying effective theory, and in particular, the best fit models of \cite{DESI:2024mwx} which have $w < -1$ for a significant portion of the evolution may be difficult to realize in a consistent effective field theory. 

\begin{figure}
    \centering
    \includegraphics[width=\textwidth]{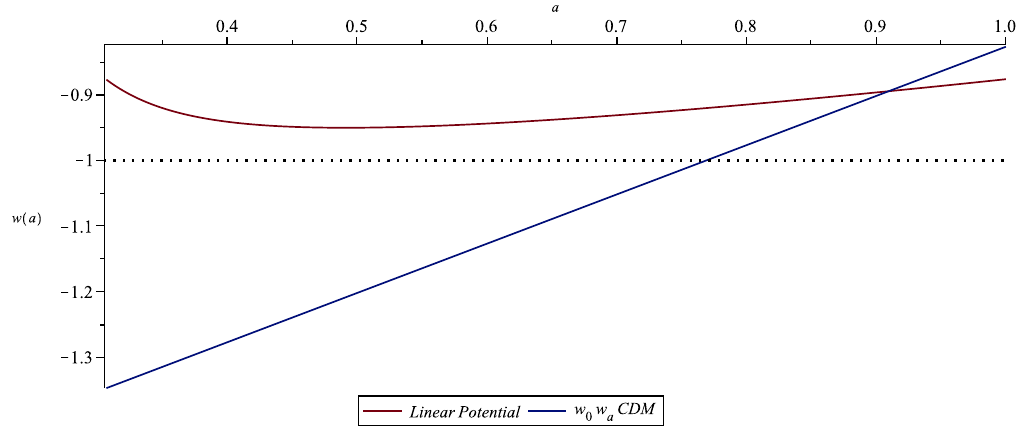}
    \caption{ Equation of state parameter $w$ versus scale factor $a$ for our best fit linear potential model using supernova and BAO data (red) and the $w_{0} w_{a}$CDM model with parameters from \cite{DESI:2024mwx}. The dashed line shows the $\Lambda$CDM value. }
    \label{fig:EoS_Linear_SNBAO}
\end{figure}

On the other hand, our starting point is a consistent effective field theory, and while $w$ also varies with time, the constraint $w \ge -1$ is guaranteed.
In Figure \ref{fig:EoS_Linear_SNBAO}, we show the evolution of the effective equation of state parameter
\begin{equation} \label{eq:eos}
    w = \frac{\frac{1}{2} \dot{\phi}^{2} - V}{\frac{1}{2} \dot{\phi}^{2} + V}
\end{equation}
in equation (\ref{eq:eos}) as a function of scale factor $a$ for our best fit linear potential model (taking into account supernova and BAO date), over the range of $a$ for which supernova data are available. We also depict $w(a)$ for the flat $w_{0} w_{a}$CDM model (presumed linear in scale factor rather than redshift, $w(a) = w_{0} + w_{a}(1-a)$), with parameters from the recent DESI analysis \cite{DESI:2024mwx} including constraints from CMB and Pantheon+. We see that for the linear potential model, $w(a)$ is {\it not} well modeled by a linear function.

Thus, the results of this paper, that $\Lambda$CDM appears to be disfavoured among linear potential models are quite distinct from the results of  \cite{DESI:2024mwx}, since we are looking at a different (and perhaps more physical) slice of the parameter space of possible models.

\subsubsection*{Theoretical implications}

The viability of models with decreasing dark energy is particularly interesting in the context of quantum gravity models arising from string theory. Here, the gravitational effective theories with a well-understood UV completion (via the AdS/CFT correspondence) presently all have $\Lambda \le 0$. These effective field theories thus cannot reproduce $\Lambda$CDM cosmologies, but they can
have cosmological solutions with late-time acceleration due to time-dependent scalar fields. Thus, the results in this paper may be taken as providing encouragement that cosmologies based on the the best-understood gravitational effective field theories arising from string theory could provide a description of our universe.

\section*{Acknowledgements}
We thank Stefano Antonini, Petar Simidzija, and Brian Swingle for
discussion and collaboration on related topics. We thank Dillon Brout, Lukas Hergt, Gary Hinshaw, and Douglas Scott for valuable discussions. We acknowledge support
from the National Science
and Engineering Research Council of Canada (NSERC) and the Simons foundation via a
Simons Investigator Award and the “It From Qubit” collaboration grant. CW would like to acknowledge support from the KITP Graduate Fellowship Program while part of this work was completed. This research was supported in part by the National Science Foundation under Grant No. NSF PHY-1748958. 

\appendix

\section{Reproducing the best fit $\Lambda$CDM scale factor with various $\Omega_M$}

According to our results above, even a completely precise knowledge of $a(t)$ still leaves is with a one-parameter family of possible potentials, where the parameter is $\Omega_M$. To illustrate this point, we suppose in this section, that the scale factor is precisely the one arising from a $\Lambda$CDM model with the parameters  ($\Omega_M^0 = 0.334$, $H_0 = 73.6 km/s/Mpc$) that provide a best fit to the Pantheon+ supernova redshift vs brightness data, and deduce the scalar potentials that would reproduce this same scale factor for smaller values of $\Omega_M$.

The $\Lambda$CDM curve $a(t; \Omega_M^0, H_0)$ corresponding to our choice of parameters is determined via the Friedmann equation in the $\Lambda$CDM model by
\be
\label{LCDM}
\dot{a} = H_0 \sqrt{(1 -\Omega_M^0)a^2 + {\Omega_M^0 \over a}} \; .
\ee
We can integrate to find $a(t)$. 

Now we would like to understand the $V(\phi)$ required to reproduce this same $a(t)$ for a model with some smaller $\Omega_M$. Making use of the rescaled variables defined above, the equation (\ref{LCDM}) together with (\ref{phisol1}) and (\ref{V3a}) gives
\beas
\frac{da}{ds} &=& \sqrt{(1 -\Omega_M^0)a^2 + {\Omega_M^0 \over a}} \cr
{d \phi \over ds} &=& {1 \over a^{3 \over 2}} \sqrt{\Omega_M^0 - \Omega_M} \cr
\tilde{V} &=& {1 \over 2 a^3} (\Omega_M^0 - \Omega_M) + (1 - \Omega_M^0)
\eeas
Dividing the second equation by the first, we get a differential equation for $\tilde{\phi}$ in terms of $a$. We have finally
\beas
\phi(a) &=& \int_a^1 d \hat{a} {\sqrt{ \Omega_M^0 - \Omega_M \over (1 - \Omega_M^0) \hat{a}^5 + \Omega_M^0 \hat{a}^2}} \cr
\tilde{V} &=& {1 \over 2 a^3} (\Omega_M^0 - \Omega_M) + (1 - \Omega_M^0).
\eeas
These two equations determine $\tilde{V}(\phi)$ parametrically. 

In Figure \ref{fig:Vplots}, we plot the deduced potentials for $\Omega_M = 0.3,0.25,0.2,0.15$ and $0.1$. We see that for lower values of $\Omega_M$, a larger amount of scalar evolution is required to produce the same $a(t)$. Even for the slightly lower value $\Omega_M = 0.3$, the scalar potential would have descended from a value 33 percent more than its present value since $z=2$ while for $\Omega_M < 0.286$ the scalar potential reproducing $\Lambda$CDM must have descended from more than twice its present value during the time frame for which supernova observations are available. 

The particular shape of the potential showing up in the examples of Figure \ref{fig:Vplots} is related to the starting assumption that the actual scale factor is precisely that of a $\Lambda$CDM model. We will see that slightly different scale factors can give substantially different potentials. 

\section{Non-relativistic matter from oscillating scalar via the WKB approximation}
\label{sec:oscillation}

In this section, we explain why models with large $V_2$ and significant oscillation in the scalar field potential energy can provide scale factor evolution very similar to a $\Lambda$CDM model.

To begine, we note that the limit $\sqrt{V_2}/H_0 \gg 1$ (in units with $8 \pi G/3 = 1$) is a limit where the scalar field equation can be well-approximated by a WKB-like approximation. Specifically, after a redefinition
\be
\phi = \psi - {V_1 \over V_2} \; , 
\ee
we can write the scalar field equation as
\be
{d^2 \psi \over dw^2}  = - p^2(w) \psi 
\ee
where 
\be
a^3 {d \over dt} = {d \over dw}
\ee
and
\be
p(w) = \sqrt{V_2} a^3(w) \; .
\ee
This is exactly the Schr\"odinger equation, where $p(w)$ is the classical momentum. This is large provided that $V_2$ is large, so we can approximate the solution by the standard WKB expression
\be
\psi(w) = {C \over \sqrt{p(w)}} e^{\pm i \int p(w) dw} \; .
\ee
Rewriting in terms of the original variables and taking a general real combination of the solutions, we get
\be
\phi(t) = -{V_1 \over V_2} + {C \over a^{3 \over 2}(t)} \cos(\sqrt{V_2} t + \varphi)
\ee
Plugging back into the Friedmann equation, the contribution to the energy density from the scalar field becomes
\be
{1 \over 2} \dot{\phi}^2 + V(\phi) = \left(V_0 - {1 \over 2} {V_1^2 \over V_2}\right) + {V_2 C^2 \over 2} {1 \over a^3} + \dots
\ee
where the omitted terms are suppressed by powers of $1/\sqrt{V_2}$  (or $H_0/\sqrt{V_2}$ before rescaling) relative to the terms present. The leading terms here behave exactly as a cosmological constant $V_0 - V_1^2 / (2 V_2)$ plus non-relativistic matter density. Thus, to an approximation that becomes increasingly good for larger $V_2$, we can reproduce the dynamics of a $\Lambda$CDM model by replacing some of the matter with an oscillating scalar field. 

That oscillating scalars in a quadratic potential can act like non-relativistic matter was explained originally in \cite{turner1983coherent}, and studied in many later works. From the quantum field theory point of view, such oscillating fields can be understood as a coherent state of a large density of the massive scalar particles associated with this scalar. The energy density of these particles (or the oscillating field) will gravitationally interact with other matter and radiation and develop inhomogeneities as with ordinary matter, so provides an interesting candidate for dark matter (see \cite{urena2019brief} for a recent review). 

However, in order for the scalar to contribute to the background evolution as we have described or to act as dark matter, it is essential that the associated scalar particles are nearly stable; this requires extremely small couplings to other fields and may be more natural with a pseudoscalar axion-like particle. 

\bibliographystyle{JHEP}
\bibliography{refs}

\end{document}